\documentclass{PoS}
\usepackage{amsmath}
\usepackage{txfonts}
\usepackage{subfigure}
\usepackage{ulem}

\newcommand{\vect}[1]{\vec{#1}}

\newcommand{\ks}[1]{#1 \!\!\! \slash }

\newcommand{\ie}{{\it i.e.}}
\newcommand{\eg}{{\it e.g.}}
\newcommand{\etal}{{\it et al.}}

\newcommand{\cf}[1]{{Fig.~\ref{#1}}}

\newcommand{\beq}[1]{
\begin{equation}\label{#1}}
\newcommand{\eeq}{\end{equation}}
\newcommand{\bea}[1]{
\begin{eqnarray}\label{#1}}
\newcommand{\eea}{\end{eqnarray}}
\usepackage{lineno}

\newcommand{\out}{\raise-3pt\hbox{\scriptsize    out}}

\title{Hard exclusive processes in the backward region}

\ShortTitle{Hard exclusive processes in the backward region}

\author{\speaker{J.P. Lansberg}\\
       IPNO, Universit\'e Paris-Sud 11, CNRS/IN2P3, F-91406, Orsay, France\thanks{Permanent address at IPNO} \\
Centre de Physique Th\'eorique, \'Ecole polytechnique, CNRS, F-91128, Palaiseau, France\\
E-mail: \email{Jean-Philippe.Lansberg@in2p3.fr}}

\author{B. Pire\\
       Centre de Physique Th\'eorique, \'Ecole polytechnique, CNRS, F-91128, Palaiseau, France\\
E-mail: \email{Bernard.Pire@cpht.polytechnique.fr}
}

\author{L. Szymanowski\\
       Soltan Institute for Nuclear Studies, Warsaw, Poland\\
E-mail: \email{Lech.Szymanowski@fuw.edu.pl}
}

\abstract{We review the potentialities offered by the study of backward exclusive processes in  a new scaling regime, 
i.e. involving a large --timelike or spacelike-- $Q^2$ photon and a baryonic exchange in the $t$-channel. We recall the concept of 
Transition Distribution Amplitudes (TDAs) containing unique information on the hadron structure, then discuss 
how they enter the description of processes such as backward electroproduction of a pion, antiproton-proton 
annihilations into a dilepton + meson as well as into $J/\psi$ + meson. We then discuss first phenomenological studies
for processes that are being analysed at JLAB and HERMES or that will be measured by Panda at GSI-FAIR. 
Finally we present outlooks for their theoretical studies based on approaches such as the pion-cloud model.}

\FullConference{35th International Conference of High Energy Physics - ICHEP2010,\\
		July 22-28, 2010\\
		Paris France}

\begin{document}

\section{Introduction}\label{aba:sec1}

According to a  well-established framework \cite{Muller}, the Bjorken limit of 
near {\it forward} exclusive reactions with a far off-shell photon allows the factorisation of the corresponding leading-twist
amplitudes into a perturbatively calculable sub-process at the quark and gluon level and 
non-perturbative hadronic matrix elements of light-cone non-local operators, in this case the Generalised Parton 
Distributions (GPDs). These are new QCD objects carrying much information on the hadronic structure.
 Besides, a further generalisation of the GPD concept has been proposed \cite{pioneer} in cases where the initial and final states are 
different hadronic states.
When these new hadronic objects are defined  through a three-quark operator (baryon-to-meson or baryon-to-photon 
transition), 
we call them {\it baryonic} Transition Distribution Amplitudes (TDAs).

In~\cite{Pire:2005ax,Pire:2005mt,Lansberg:2007ec}, we have introduced the framework to study
{\it backward} pion electroproduction, $\gamma^\star(q) N(p_1)  \to N'(p_2) \pi(p_\pi),$
on a proton (or neutron) target, in the Bjorken regime ($q^2$ large and $q^2/(2 p_1.q)$ fixed) 
in terms of a factorised amplitude (see \cf{fig:fact} (a)) where a hard part is convoluted with  the aforementioned TDAs. The same framework 
can also be applied to the reaction,
$N(p_1) \bar N (p_2) \to \gamma^\star(q) \pi(p_\pi)$,
in the  near forward region~\cite{Pire:2004ie,Lansberg:2007se,Lutz:2009ff} (see \cf{fig:fact} (b)) .
In the following, we first recall the definition of the 
TDAs in the exemplary  $p \to \pi^0$ case, and then discuss 
how they enter the description of processes such as backward electroproduction of a pion,
antiproton-proton annihilations into a dilepton + meson as well as into $J/\psi$ + meson. 
Finally, we present some outlooks.

\begin{figure}[htb!]
\centering{
\subfigure[\scriptsize $\pi^0$ electroproduction at small $u$]{\includegraphics[width=0.3\textwidth,clip=true]{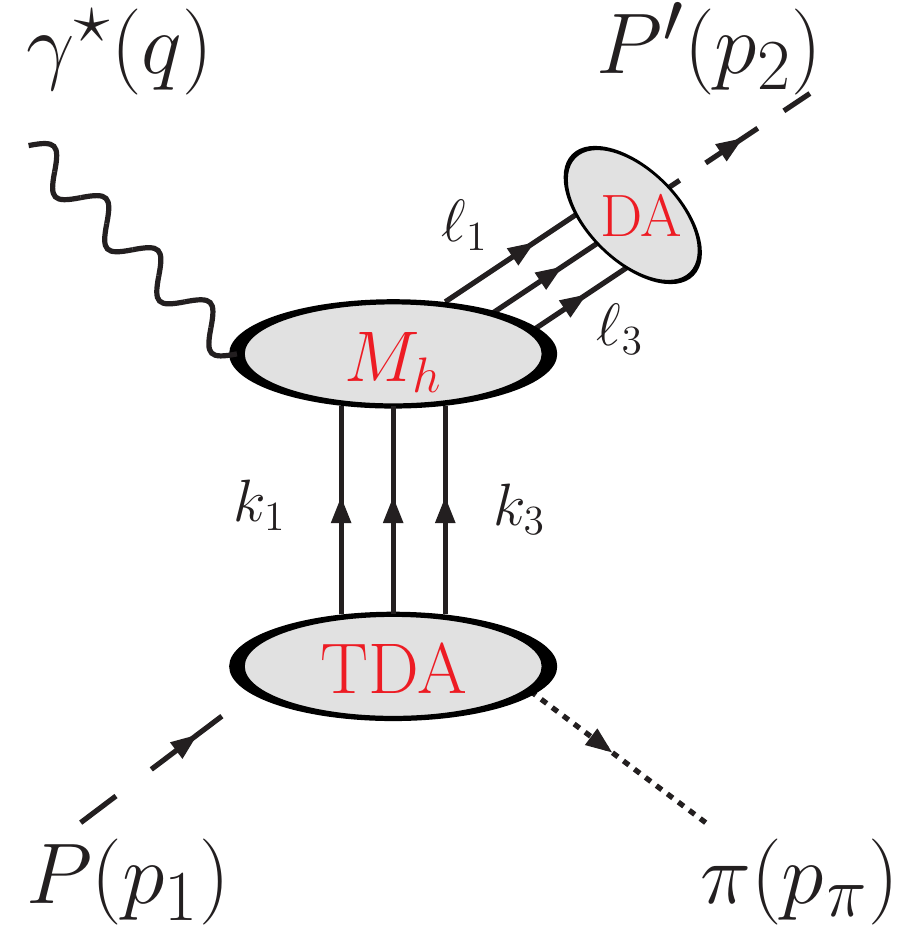}}
\subfigure[\scriptsize $p\bar p\to \gamma^\star \pi^0$ at small $t=(p_p-p_{\pi^0})^2$]{\includegraphics[width=0.3\textwidth,clip=true]{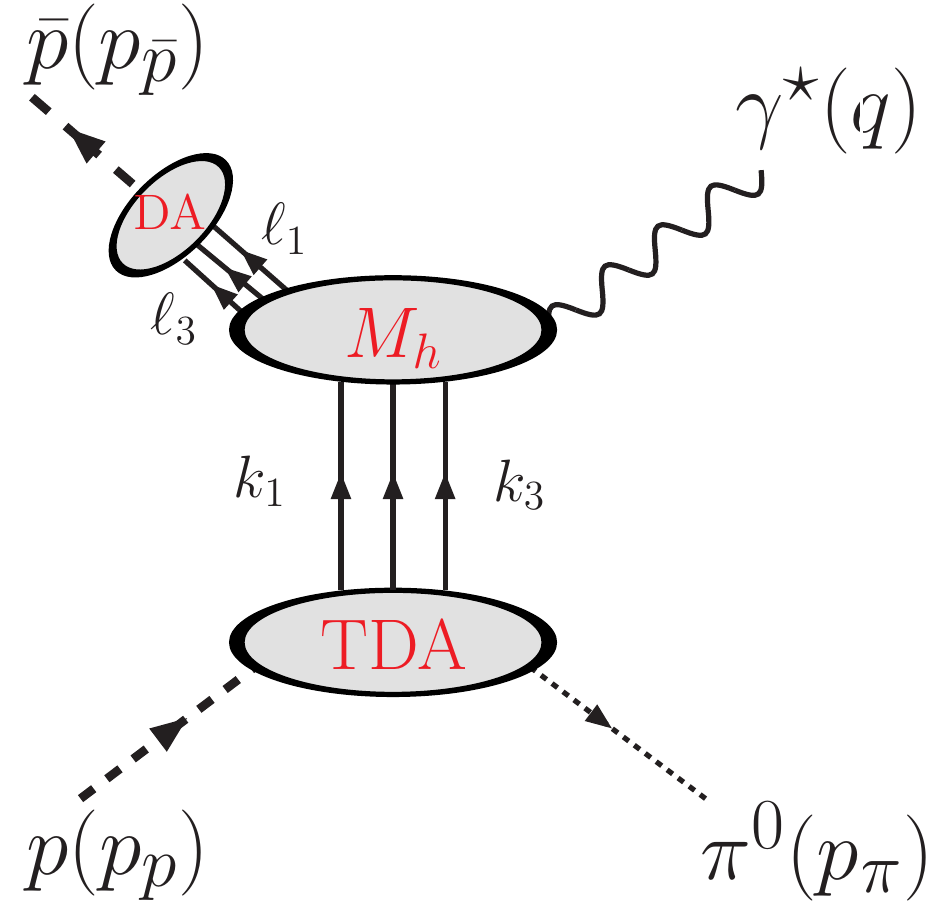}}
\subfigure[\scriptsize$p\bar p\to J/\psi \pi^0$ at small $t=(p_p-p_{\pi^0})^2$ ]{\includegraphics[width=0.3\textwidth,clip=true]{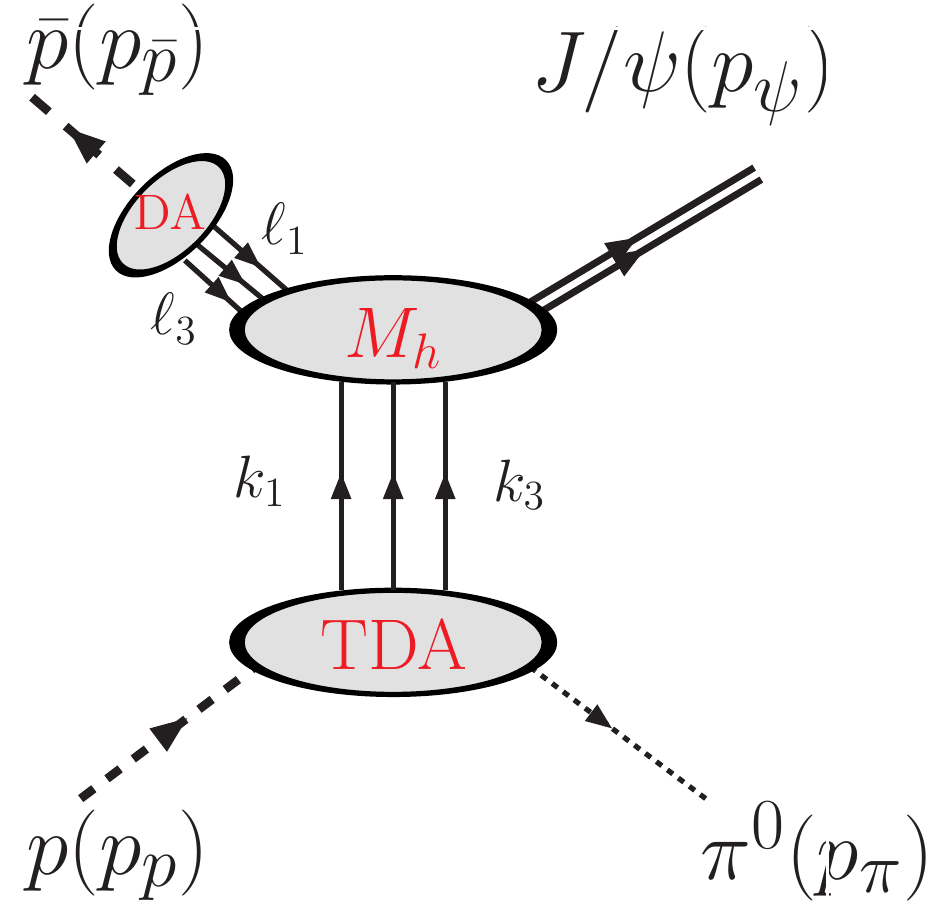}}
}
\caption{Illustration of the factorisation for three exclusive reactions involving the TDAs.}
\label{fig:fact}
\end{figure}

\section{Definition of baryonic TDAs}
The leading-twist TDAs for the $p \to \pi^0$ transition, $ V^{p\pi^0}_{i}\!\!(x_i,\xi, \Delta^2)$, 
$A^{p\pi^0}_{i}\!\!(x_i,\xi, \Delta^2)$ and 
$T^{p\pi^0}_{i}\!\!(x_i,\xi, \Delta^2)$  are defined~\footnote{The present definitions differ from these 
of~\cite{Pire:2005ax} by constant multiplicative factors and by the definition of $\sigma^{\mu\nu}$.}  
 as :
\bea{TDA}
&& 4 {\cal F}\Big(\langle     \pi^0(p_\pi)|\, \epsilon^{ijk}u^{i}_{\alpha}(z_1 n) 
u^{j}_{\beta}(z_2 n)d^{k}_{\gamma}(z_3 n)
\,|P(p_1) \rangle \Big)  
=   i\frac{f_N}{f_\pi}\Big[ V^{p\pi^0}_{1} (\ks p C)_{\alpha\beta}(N^+)_{\gamma}+A^{p\pi^0}_{1} (\ks p\gamma^5 C)_{\alpha\beta}(\gamma^5 N^+)_{\gamma}
\nonumber\\ 
&&
+
T^{p\pi^0}_{1} (\sigma_{p\mu} C)_{\alpha\beta}(\gamma^\mu N^+)_{\gamma} 
+ M^{-1}V^{p\pi^0}_{2} 
(\ks p C)_{\alpha\beta}(\ks \Delta\!_T N^+)_{\gamma} +M^{-1}
A^{p\pi^0}_{2}(\ks p \gamma^5 C)_{\alpha\beta}(\gamma^5\ks \Delta\!_T N^+)_{\gamma}
\\
&& 
+ M^{-1}T^{p\pi^0}_{2} ( \sigma_{p\Delta_T} C)_{\alpha\beta}(N^+)_{\gamma}
+ M^{-1}T^{p\pi^0}_{3} ( \sigma_{p\mu} C)_{\alpha\beta}(\sigma^{\mu\Delta_T}
N^+)_{\gamma} + M^{-2}T^{p\pi^0}_{4} (\sigma_{p \Delta_T} C)_{\alpha\beta}
(\ks \Delta\!_T N^+)_{\gamma}\;\Big], \nonumber
\eea
where%
\footnote{We  use the notation $\displaystyle {\cal F}\equiv (p.n)^3\int^{\infty}_{-\infty} \Pi_j dz_j/(2\pi)^3 e^{i\Sigma_k x_k z_k p.n}$.
The momenta of the process $\gamma^\star P \to P' \pi $ are defined as in \cf{fig:fact} (a).
The $z$-axis is chosen along the initial nucleon and the virtual photon  momenta
and the $x-z$ plane is identified 
with the collision or hadronic plane. Then, we define the 
light-cone vectors $p$ and $n$ ($p^2$=$n^2$=0) 
such that $2~p.n=1$, as well as $P=\frac{1}{2} (p_1+p_\pi)$, $\Delta=p_\pi -p_1$ and its 
transverse component $\Delta_T$ ($\Delta_T.\Delta_T=\Delta_T^2<0$). From these, we define  $\xi$ in an usual way as $\xi=-\frac{\Delta.n}{2P.n}$.
Expressing the momenta of the particles through their  
Sudakov decomposition and, keeping the first-order corrections in the masses and $\Delta_T^2$, we have:
$
p_1= (1+\xi) p + \frac{M^2}{1+\xi}n, q\simeq- 2 \xi \Big(1+ \frac{(\Delta_T^2-M^2)}{Q^2}\Big)  p + \frac{Q^2}{2\xi\Big(1+ \textstyle 
\frac{(\Delta_T^2-M^2)}{Q^2}\Big)} n,p_\pi= (1-\xi) p +\frac{m_\pi^2-\Delta_T^2}{1-\xi}n+ \Delta_T,
\Delta= - 2 \xi p +\Big[\frac{m_\pi^2-\Delta_T^2}{1-\xi}- \frac{M^2}{1+\xi}\Big]n
+ \Delta_T, p_2\simeq- 2 \xi \frac{(\Delta_T^2-M^2)}{Q^2} p+ \Big[\frac{Q^2}{2\xi\Big(1+ \textstyle \frac{(\Delta_T^2-M^2)}{Q^2}\Big)} -
\frac{m_\pi^2-\Delta_T^2}{1-\xi}+ \frac{M^2}{1+\xi}\Big]n - \Delta_T.$}
$\sigma^{\mu\nu}= 1/2[\gamma^\mu, \gamma^\nu]$ with $\sigma^{p \mu} = p_\nu \sigma^{\nu\mu}$,..., $C$ is the charge 
conjugation matrix 
and $N^+$ is the large component of the nucleon spinor 
($N=(\ks n \ks p + \ks p \ks n) N = N^-+N^+$
with $N^+\sim \sqrt{p_1^+}$ and $N^-\sim \sqrt{1/p_1^+}$). The TDAs $V_i$, $A_i$ and $T_i$ are
dimensionless. Note that the first three 
terms in (\ref{TDA}) are the only ones surviving the limit $\Delta_T \to 0$.

Baryonic TDAs are matrix elements of the same operator that appears in baryonic 
Distribution Amplitudes (DAs). The 
known evolution equations of this operator lead to derive evolution equations which have different 
forms in different regions ; one defines one ERBL-like and two types of DGLAP-like regions
much in the same spirit as in the GPD case, so that the evolution equations in momentum space depend on 
the signs of the quark momentum fractions $x_{i}$.
As for DAs, an asymptotic solution for this evolution equation exists but the phenomenological study of
electromagnetic form factors leads us to strongly doubt that it is of any phenomenological relevance. 
 We thus do not propose to take an asymptotic TDA as a realistic input for phenomenology.

On the other hand, there exists an interesting soft limit \cite{Lansberg:2007ec} when the emerging pion momentum is small, which allows one
to relate proton $\to$
pion TDAs to proton DAs. The well-known soft pion theorems indeed gives:
\begin{equation}
\langle \pi^a(p_\pi)  |{\cal O}| P(p_1,s_1)\rangle \to -\frac{i}{f_\pi} \langle 0  | [ Q^a_5,  {\cal O}]  | P(p,s) \rangle \\ \nonumber
\nonumber
\label{eq:soft-theorem}
\end{equation}
when  $|\vect p_\pi|\to 0$. 
One then gets very simple relations
between the nucleon DAs $A^p$, $V^p$ and $T^p$ on the one hand and the $p\to \pi^0$ TDAs $V^{p\pi^0}_{1}$, 
$A^{p\pi^0}_{1}$ and $T^{p\pi^0}_{1}$ on the other hand.
 To get 
theoretical insights on the TDAs away from $|\vect p_\pi|\to 0$ ($\xi \to 1$ if one neglects $m_\pi$), 
one has to resort to general arguments such as their spectral representation~\cite{Pire:2010if}
or to use models which successfully describe other hadronic observables~\cite{Pasquini:2009ki}. Let us finally mention 
that, for the time being, there exists no modelling of the proton to photon TDAs, entering the description of
backward DVCS~\cite{Lansberg:2006uh}.

\section{Processes involving the TDAs and the experimental situation}

As we have mentioned above, $p\to \pi$ baryonic TDAs appear in
the description of backward electroproduction of a pion on a proton target (see \cf{fig:fact} (a)). In terms of
angles, in the $\gamma^\star p$ center of momentum (CM) frame, the angle between
the $\gamma^\star$ and the pion, $\theta^\star_\pi$, is close to 180$^\circ$.
We then have $|u|\ll s$ and $t\simeq -(s+Q^2)$, in contrast to the fixed angle regime 
$u\simeq t\simeq -(s+Q^2)/2$ ($\theta^\star_\pi\simeq 90^\circ$) and the forward (GPD) one
$|t|\ll s$ and $u\simeq -(s+Q^2)$ ($\theta^\star_\pi\simeq 0^\circ$).

The TDAs appear also in similar electroproduction processes such 
as $e p \to  e\;(p,\Delta^+) \; (\eta,\rho^0)$,  $e p \to  e\;(n,\Delta) \; (\pi^+,\rho^+)$, $e p \to  e\;\Delta^{++}\; (\pi^-,\rho^-)$. 
These processes have already been analysed, at
backward angles, at JLab in the resonance region, \ie~$\sqrt{s_{\gamma^\star p}}=W<1.8$ GeV, in order to study the baryonic 
transition form factors in the $\pi$ channel~\cite{Park:2007tn} or in the $\eta$ 
channel~\cite{Armstrong:1998wg,Denizli:2007tq}. 
Data is being extracted in some channels above the 
resonance region. The number of events seems large enough to expect to get cross section measurements for 
$|\Delta^2_T|<1$ GeV$^2$, which is the region described in terms of TDAs. 
Hermes analysis~\cite{Airapetian:2007an} for forward electroproduction may  also be extended to larger values of  
$-t$.  It has to be noted though, that present studies are limited to $Q^2$ of order a few GeV$^2$, 
which gives no guarantee to reach the TDA regime yet.
Higher-$Q^2$ data may be obtained at JLab-12 GeV and in muoproduction at Compass within the next few years.

Crossed reactions in proton-antiproton annihilation (\eg~with PANDA at GSI-FAIR~\cite{Lutz:2009ff}), 
with time-like photons (\ie~di-leptons) and a pion (see \cf{fig:fact} (b)) also involve TDAs
both for small $t=(p_p-p_{\pi^0})^2$ and $u=(p_{\bar p}-p_{\pi^0})^2$. In the latter case, the TDAs
for a transition between the anti-proton and the pion are probed. One can also  study similar
reactions with other mesons than a pion, \eg~$\bar p p \to  \gamma^\star\; (\eta,\rho^0) $, or on a different target
than proton $\bar p N \to \gamma^\star\pi$. Finally, one may consider 
$J/\psi$ production in association  with a pion  $\bar p p \to  \psi\; \pi^0 $  (see \cf{fig:fact} (c)),
which involves the {\it same} TDAs as  $\bar p p \to   \gamma^\star\; \pi^0 $ and the backward electroproduction of
a $\pi^0$. These studies will serve as  very strong tests of the universality of the TDAs in different processes. 

The first application of baryonic TDAs was centered on backward electroproduction of a 
pion~\cite{Lansberg:2007ec}. In that case, the hard contribution, which consists in the 
scattering of the hard photon with three quarks, is known at the leading order.
Extrapolating the limiting value of the TDAs obtained from the soft pion theorems to the large-$\xi$ 
region, we have obtained a first evaluation of the unpolarised cross section for 
backward electroproduction. This estimate, which is unfortunately reliable only in a very
restricted kinematical domain (large-$\xi$), shows an interesting sensitivity to the underlying 
model for the proton DAs. This study has been extended to hard exclusive production of a $\gamma^\star \pi^0$ pair in $\bar p p $ 
annihilation at GSI-FAIR~\cite{Lansberg:2007se}.

More work is needed before being able to proceed to  quantitative {\it comparisons}  between different
TDA {\it models} as well as between theory and experiments. For the time being, model
independent analyses -- looking for scaling or dominance of $\sigma_T$ -- sound more expedient. 
In this context, we have argued in~\cite{Lansberg:2010pf} that the study of target transverse-spin asymmetry 
could be used as a test of the 
dominance of a hard parton-induced scattering in the backward region at large $Q^2$ rather than that of a soft baryon exchange 
in the $u$-channel. Such a reaction would only generate phases through final state interactions, 
expected to decrease for $W^2 \gg (M+m_\pi)^2$ and large $Q^2$; the SSA would then be vanishing. On the contrary, 
in scatterings at the parton level, one expects an imaginary part to develop and to 
generate a SSA independently of whether $W^2$ and $Q^2$ are large or not.

\section{Conclusions}

Backward hard-exclusive reactions thus open a new window in the understanding
of hadronic physics in the framework of the
collinear-factorisation approach of QCD. Of course, the
most important and most difficult problem to solve, in
order to extract reliable and precise information on the baryon-to-meson 
transition  from an incomplete set
of observables such as cross sections and asymmetries, is
to develop a realistic model for the TDAs. This is the
subject of nonperturbative studies such as, e.g., lattice
simulations. Three approaches --the soft pion theorem, the pion-cloud model
and the spectral representation-- are being explored and first 
cross-section evaluations in the whole kinematical domain covered
by the TDA factorisation should be available soon.

We are also hopeful that the 12 GeV JLab upgrade and the start-up of
GSI-FAIR will bring us the necessary experimental information
to test the model-independent predictions of the TDA factorisation
and then to check specific predictions from different TDA models, potentially
connected to more fundamental quantity of the hadronic realm.

\section*{Acknowledgments}
We thank S.J. Brodsky, G. Huber, V. Kubarovsky, K.J. Park, B.~Pasquini, K.~Semenov-Tian-Shansky, P. Stoler and S. Wallon 
for useful and motivating discussions.
This work is partly supported by the ANR contract BLAN07-2-191986. L.Sz. acknowledges the support 
by the Polish Grant N202 249235.

\end{document}